\documentclass[journal]{IEEEtranTIE}
\usepackage{graphicx}
\usepackage{cite}
\usepackage{picinpar}
\usepackage{amsmath}

\usepackage{url}
\usepackage{flushend}
\usepackage[utf8]{inputenc}
\usepackage{colortbl}
\usepackage{soul}
\usepackage{multirow}
\usepackage{pifont}
\usepackage{color}
\usepackage{alltt}
\usepackage{amsthm}

\usepackage[hidelinks]{hyperref}
\usepackage{enumerate}
\usepackage{siunitx}
\usepackage{booktabs} 
\usepackage{url}
\usepackage{hyperref}
\usepackage{cleveref}
\crefname{figure}{fig}{figures}
\Crefname{figure}{Fig}{Figures}
\hypersetup{hidelinks}
\usepackage{amssymb}
\usepackage{epstopdf}
\usepackage{pbox}
\usepackage{verbatim}
\usepackage{bm}

\usepackage{siunitx,booktabs}
\begin{document}
\title{Adaptive Lyapunov-constrained MPC for fault-tolerant AUV trajectory tracking}

\author{Haolin Liu, Shiliang Zhang, Xiaohui Zhang, Shangbin Jiao$^*$,  Xuehui Ma, Ting Shang, Yan Yan, Wenqi Bai, Youmin Zhang

\thanks{$*$Corresponding author.\\This work was supported  by the National Natural Science Foundation of China via the grant No. 62371388, 62127809, and 62103082, and by Scientists \& Engineers Team (No.23KGDW0011 and 2024QCY-KXJ-014), and by the Qin Chuangyuan Innovation Platform (No.23TSPT0002). Haolin Liu, Shangbin Jiao, Xiaohui Zhang,  Xuehui Ma, Yan Yan, and Wenqi Bai  are with Xi'an University of Technology, Xi'an, China (haolinliu@xaut.edu.cn, jiaoshangbin@xaut.edu.cn, xhzhang@xaut.edu.cn, xuehui.yx@gmail.com, yan.y@xaut.edu.cn, bayouenqy@outlook.com). Shiliang Zhang is with University of Oslo, Norway (shilianz@ifi.uio.no). Youmin Zhang is with Concordia University, Montreal, Canada (youmin.zhang@concordia.ca).}
}

\maketitle
	
\begin{abstract}
Autonomous underwater vehicles (AUVs) are subject to various sources of faults during their missions, which challenges AUV control and operation in real environments. This paper addresses fault-tolerant trajectory tracking of autonomous underwater vehicles (AUVs) under thruster failures. We propose an adaptive Lyapunov-constrained model predictive control (LMPC) that guarantees stable trajectory tracking when the AUV switches between fault and normal modes. Particularly, we model different AUV thruster faults and build online failure identification based on Bayesian approach. This facilitates a soft switch between AUV status, and the identified and updated AUV failure model feeds LMPC controller for the control law derivation. The Lyapunov constrain in LMPC ensures that the trajectory tracking control remains stable during AUV status shifts, thus mitigating severe and fatal fluctuations when an AUV thruster occurs or recovers. We conduct numerical simulations on a four-thruster planar AUV using the proposed approach. The results demonstrate smooth transitions between thruster failure types and low trajectory tracking errors compared with the benchmark adaptive MPC and backstepping control with rapid failure identification and failure accommodation during the trajectory tracking.
\end{abstract}



\definecolor{limegreen}{rgb}{0.2, 0.8, 0.2}
\definecolor{forestgreen}{rgb}{0.13, 0.55, 0.13}
\definecolor{greenhtml}{rgb}{0.0, 0.5, 0.0}

\section{Introduction}
\IEEEPARstart{A}{utonomous} underwater vehicles (AUVs) enable ocean observation, inspection, and intervention~\cite{11124293,10901967,bai2024long,yang2024hardware}. Among AUV propulsion solutions, propeller thrusters dominate practical platforms because of its advantages in efficiency and robustness~\cite{Fossen2021,Ma_2025}. Reliable trajectory tracking in missions is a core capability for thrusters~\cite{Chu2024,Chen_2021}. In field operations, the control of thruster is subject to faults that arise from blockage or entanglement, blade breakage, mechanical looseness, and motor or gear failures~\cite{Liu_2023,liu2025adaptive}. Reported statistics indicate that thrusters account for the majority of actuator faults in AUVs~\cite{Liu_2023,Liu_2024}. Such faults can degrade tracking accuracy and and lead to failed missions and economic loss~\cite{Li_2023}. Therefore, the control of thrusters under fault conditions, a.k.a. fault-tolerant control, is of critical importance particularly in real world AUV missions.

The objective of fault-tolerant control (FTC) is to maintain stability and tracking accuracy under loss of effectiveness (LOE) and misalignment while respecting physical constraints. Recent FTC studies span passive and active schemes, diagnosis-aware reconfiguration, and robust or adaptive control~\cite{10551447,Wang2024,Liu_2023,10591250}. Multiple-model (MM) FTC has gained traction because it encodes a library of failure scenarios and coordinates controllers accordingly. However, many MM strategies rely on hard switching. Classical MM adaptive control and supervisory switching theory show that hard switching can induce transients unless hysteresis or dwell-time logic is enforced~\cite{Narendra_2003,Hespanha}. With tight actuator limits and strong hydrodynamic couplings, these transients are often pronounced at fault instants, which motivates \emph{soft switching} based on probability-weighted blending rather than winner-take-all selection~\cite{Liu_2023,Liu2025}.

Within the controller bank, baseline laws (e.g.  PID, LQR, Backstepping Control(BSC) and Sliding Mode Control(SMC)) remain attractive for simplicity and guaranteed basic performance~\cite{Chen2023,Li2024,10055961}. Model predictive control (MPC)~\cite{zhang2017data,wang2025fine} is increasingly preferred in modern banks because it handles constraints and can integrate thrust allocation in a receding-horizon optimization~\cite{Wei_2023}. For nonlinear AUV dynamics, a finite-horizon MPC does not guarantee closed-loop stability or recursive feasibility without terminal ingredients. Lyapunov-based MPC (LMPC) addresses this gap by enforcing a \emph{first-step Lyapunov contraction} derived from a stabilizing backstepping law. This yields stability under hard constraints and an integrated formulation with thrust allocation. LMPC has been demonstrated for dynamic positioning and trajectory tracking, with distributed and robust extensions~\cite{Shen_2017,Shen2018,ScienceDirect2020,Gong2021,Gong2022,Wang2024}. In severe LOE or misalignment, however, the quadratic objective must trade off state accuracy against control effort. If the prediction model does not reflect degraded actuation, practical MPC implementations may reduce effective actuation prematurely. To mitigate this, we introduce an \emph{augmented-state LMPC}: slow force/torque biases and per-thruster LOE/misalignment parameters are estimated online as pseudo-states so that the prediction model and the allocation map track the prevailing fault~\cite{Gong2022,Gong2021,Zhang2023}.

On the inference side, we compute mode probabilities from innovation likelihoods using an Interacting Multiple Model (IMM) Bayesian recursion for Markovian switching systems~\cite{Blom_1988}. The IMM update provides a principled and recursive mechanism to fuse evidence from a bank of mode-conditioned estimators. The resulting \emph{posterior probabilities} schedule the controller bank via maximum-a-posteriori (MAP) selection or \emph{probability-weighted blending} of mode-conditioned LMPC commands, followed by projection onto the admissible input set. This soft-fusion strategy preserves the per-mode LMPC contraction and mitigates switching transients relative to hard switching~\cite{Narendra_2003,Hespanha,Shen_2018,Deshpande1973}.

\textbf{Contributions.} We propose an \emph{adaptive Lyapunov-Constrained MPC (ALMPC)} architecture with three elements: (i) a bank of mode-conditioned LMPCs sharing a first-step Lyapunov contraction; (ii) a bias-augmented prediction model that estimates slow force/torque biases and per-thruster loss-of-effectiveness/misalignment online; (iii) an IMM-style Bayesian posterior that drives probability-weighted fusion with input projection. On a four-thruster planar AUV with nominal and two fault modes, ALMPC achieves smoother transitions and lower tracking error than Adaptive MPC(AMPC) and BSC while respecting input amplitude and rate limits.

\textbf{Organization.} Section II presents the AUV model, thrust mapping, and the augmented error model. Section III develops the ALMPC design. Section IV describes the Bayesian multi-model framework and control fusion. Section V reports simulations and comparisons. Section VI concludes.

\section{PRELIMINARIES AND PROBLEM FORMULATION}\label{section:PRELIMINARIES}

\begin{figure}[t]
    \centering
    \includegraphics[width=1\columnwidth]{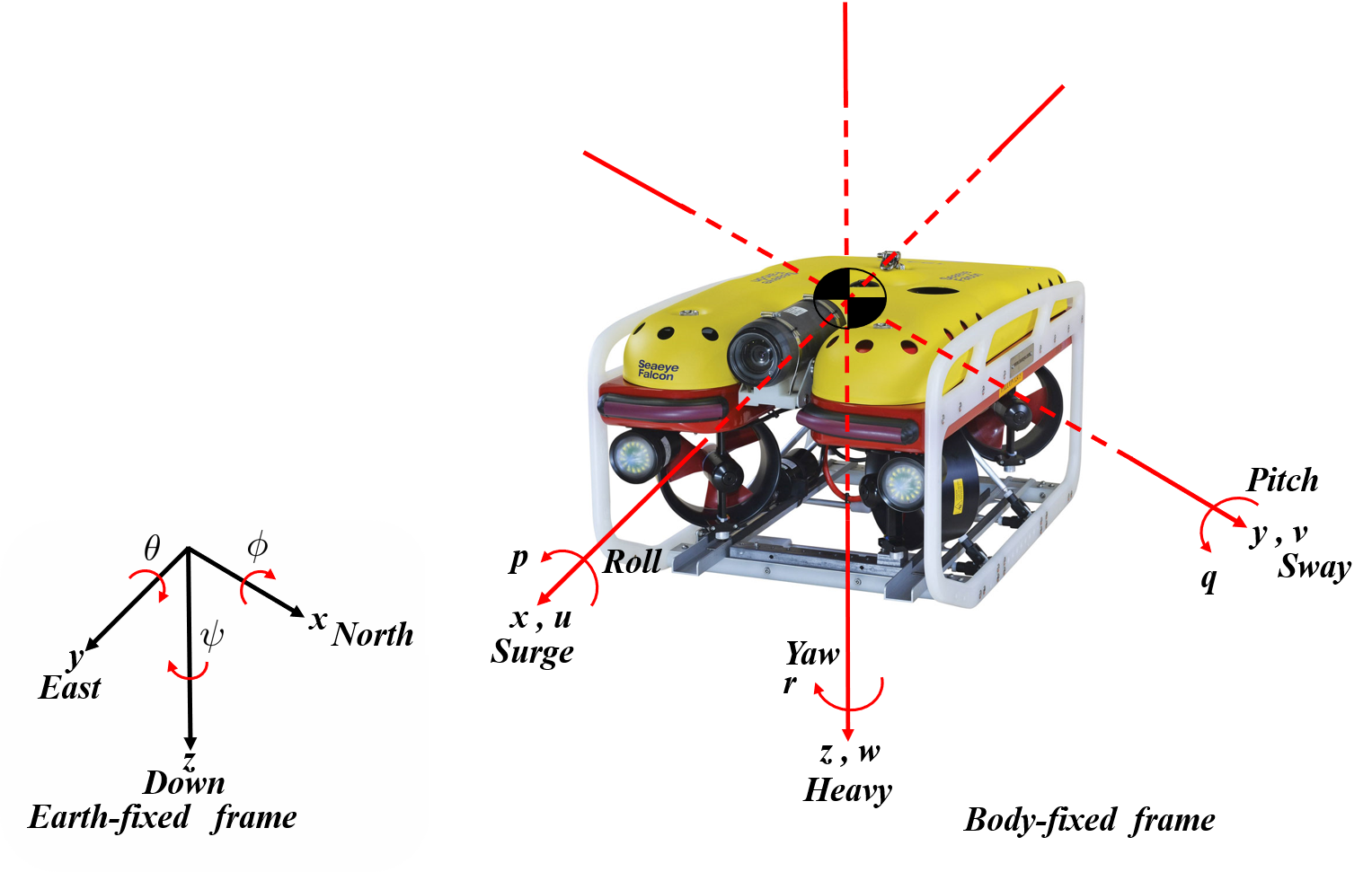}
    \caption{Earth-fixed and body-fixed frames used for the Saab Seaeye Falcon.}
    \label{fig:2}
\end{figure}

This section introduces the 3-DOF kinematics and dynamics of the Saab Seaeye Falcon AUV and formulates the augmented error model used by the controller. Owing to its thruster configuration, the vehicle is underactuated in roll and pitch; thus we restrict attention to horizontal-plane motion and trajectory tracking in the Earth-fixed frame, see \Cref{fig:2}.

\subsection{AUV Kinematics and Dynamics}
The planar kinematics are
\begin{equation}\label{eq:kinematics}
    \dot{\bm\eta} = \bm J(\psi)~\bm\nu,
\end{equation}
where $\bm\eta=[x,y,\psi]^\top$ is the Earth-fixed position and heading, $\bm\nu=[u,v,r]^\top$ is the body-fixed velocity, and
\begin{equation}\label{eq:J}
    \bm J(\psi)=\begin{bmatrix}
        \cos\psi & -\sin\psi & 0\\
        \sin\psi & \phantom{-}\cos\psi & 0\\
        0 & 0 & 1
    \end{bmatrix}=\begin{bmatrix}
        \bm R (\psi)&\bm 0\\
        0&1
    \end{bmatrix}.
\end{equation}
The 3-DOF Newton--Euler dynamics in the body frame are
\begin{equation}\label{eq:dynamics}
    \bm M~\dot{\bm\nu} + \bm C(\bm\nu)\bm\nu + \bm D(\bm\nu)\bm\nu
    \;=\; \bm\tau + \bm w,
\end{equation}
where $\bm M=\bm M^\top\!\succ0$ is the inertia matrix (including added mass), $\bm C(\bm\nu)=-\bm C^\top(\bm\nu)$ is the Coriolis--centripetal matrix, $\bm D(\bm\nu)\succ0$ is the hydrodynamic damping, $\bm\tau=[F_u,F_v,F_r]^\top$ is the generalized force/moment, and $\bm w$ collects bounded external disturbances.

Define the full state $\bm x=[\bm\eta^ \top,\bm\nu^ \top]^ \top$. The state-space form follows from \eqref{eq:kinematics}--\eqref{eq:dynamics}:
\begin{equation}\label{eq:state}
    \dot{\bm x}=
    \begin{bmatrix}
        \bm J(\psi)\bm\nu\\
        \bm M^{-1}\!\big(\bm\tau - \bm C(\bm\nu)\bm\nu - \bm D(\bm\nu)\bm\nu\big)
    \end{bmatrix}
    \;=\; \bm f(\bm x,\bm\tau).
\end{equation}

\subsection{Thruster Mapping and Fault Parameterization}
The generalized force is produced by four thrusters via the allocation model
\begin{equation}\label{eq:allocation}
    \bm\tau \;=\; \bm T(\bm\theta)~\bm\Gamma~\bm u,
\end{equation}
where $\bm u\in\mathbb{R}^{4\times 1}$ stacks the individual thruster commands, $\bm T(\bm\theta)=[\bm t_1~\bm t_2~\bm t_3~\bm t_4]\in\mathbb{R}^{3\times4}$ encodes the horizontal-plane geometry, and $\bm\Gamma=\mathrm{diag}(\gamma_1,\gamma_2,\gamma_3,\gamma_4)\in\mathbb{R}^{4\times4}$ models per-thruster effectiveness with $\gamma_i\in[0,1]$. A misalignment of thruster $i$ by angle $\theta_i$ is captured \emph{column-wise} by
\[
    \bm T(\bm\theta)=\big[\bm t_1,~\ldots,~ \bm R(\theta_i)\bm t_i,~\ldots,~\bm t_4\big],
\]
with planar rotation $\bm R(\theta_i)$. Nominal operation corresponds to $\gamma_i=1,\theta_i=0$; a blocked thruster has $\gamma_i=0$.

\subsection{Control Objective and Standing Assumptions}
The objective is to design a fault-tolerant tracking controller that  achieves high-accuracy tracking in nominal conditions, and detects and accommodates thruster faults with minimal delay, and  reallocates control effort while preserving closed-loop stability under thruster degradation. We make the following standard assumptions:
\begin{itemize}
  \item $\bm M\succ 0$, $\exists~d_{\min}>0$ s.t. $\bm D(\bm\nu)\succeq d_{\min}\bm I$ \quad
  $\forall\,\,\bm \nu=[u,\,v,\,r]^\top$ with $|u|\le \bar u,\ |v|\le \bar v,\ |r|\le \bar r$. where $\bar{\bm\nu}=[\,\bar u,\,\bar v,\,\bar r\,]^\top$ denotes the operating-range bounds on the body-frame velocities.
  \item disturbances $\bm w$ are bounded; inputs $\bm u\in[\bm u_{\min},\bm u_{\max}]$ with known rate limits; 
  \item $(\bm\Gamma,\bm\theta)$ are piecewise-constant and change at isolated instants (Sec.~\ref{sec:bayes}).
\end{itemize}
The constant $d_{\min}>0$ is a uniform damping lower bound. 
Let $\bm x_d=[~\bm\eta_d^\top,~\bm\nu_d^\top~]^\top$  denote the desired trajectory and its compatible body-frame velocity, with
\begin{equation}\label{eq:vd}
     \bm\nu_d \;=\; \bm J^{-1}(\psi_d)~\dot{\bm\eta}_d,
\end{equation}
Define the tracking error $\tilde{\bm x}=\bm x-\bm x_d$.

\subsection{Augmented Error Model}

To explicitly capture \emph{persistent, low–frequency mismatches}—such as steady currents, unmodeled hydrodynamic loads, or residual allocation errors—we introduce a slowly varying bias in the generalized-force channel and treat it as an augmented state. This bias acts like a  integral action inside MPC: it cancels constant/slow disturbances to remove steady-state offset, while remaining bounded when the mismatch disappears. We augment the generalized force with a bias $\bm d\in\mathbb{R}^{3 \times 1}$:
\begin{equation}\label{eq:tau_eff}
\begin{aligned}
    \bm\tau_{\mathrm{eff}} \;&=\; \bm T(\bm\theta)\bm\Gamma~\bm u \;+\; \bm d,\qquad \\
    \dot{\bm d}&=-\bm\Lambda \bm d,\ \ \bm\Lambda=\mathrm{diag}(\lambda_i)\succ0.
\end{aligned}
\end{equation}
Here $\bm\Lambda$ sets the bias dynamics: each $\lambda_i>0$ is a designer-chosen decay rate (units s$^{-1}$), with time constant $\tau_i=1/\lambda_i$. Small $\lambda_i$ lets $\bm d$ track slow drifts; larger $\lambda_i$ makes $\bm d$ decay faster. In practice, choose $\tau_i$ on the order of several sampling periods but slower than the dominant closed-loop bandwidth.

Using \eqref{eq:state}, \eqref{eq:allocation}, \eqref{eq:vd}, and \eqref{eq:tau_eff}, the augmented error dynamics become
\begin{equation}\label{eq:aug_error}
    \dot{\tilde{\bm x}} =
    \begin{bmatrix}
        \bm J(\psi)\bm\nu - \dot{\bm\eta}_d\\[2pt]
        \bm M^{-1}\!\big(\bm\tau_{\mathrm{eff}} - \bm C(\bm\nu)\bm\nu - \bm D(\bm\nu)\bm\nu\big)- \dot{\bm\nu}_d
    \end{bmatrix}
    \!+\; \bm w,
\end{equation}
where the residual $\bm w$ collects fast, zero-mean disturbances not captured by $\bm d$. 
and the augmented state is $\tilde{\bm x}_{\mathrm{aug}}=\big[\;\tilde{\bm x}^\top\ \ \bm d^\top\big]^\top$. The bias dynamics in \eqref{eq:tau_eff} act as an integral action in the generalized-force domain: constant or slowly varying disturbances are absorbed by $\bm d$, enabling offset-free tracking while guaranteeing that $\bm d$ decays when no longer needed. This augmented model will be used in the LMPC design (Sec.~\ref{sec:LMPC}) and in the multi-model estimator (Sec.~\ref{sec:bayes}).


\section{Adaptive Lyapunov-Based MPC  Design}\label{sec:LMPC}
Building on the formulation in Sec.~\ref{section:PRELIMINARIES}, we design a fault-tolerant controller as a \emph{Lyapunov-Constrained MPC} posed on the 6-state tracking-error model.
The augmentation captures slow-varying disturbances and model mismatch, preventing steady-state offsets under thruster degradation. A model-consistent online reconfiguration of the thrust effectiveness is embedded, so that constraints and allocation remain feasible under faults. Stability is enforced via a Lyapunov contraction constraint that each MPC must satisfy.
\subsection{Architecture Overview}
\begin{figure}[t]
    \centering
    \includegraphics[width=1\columnwidth]{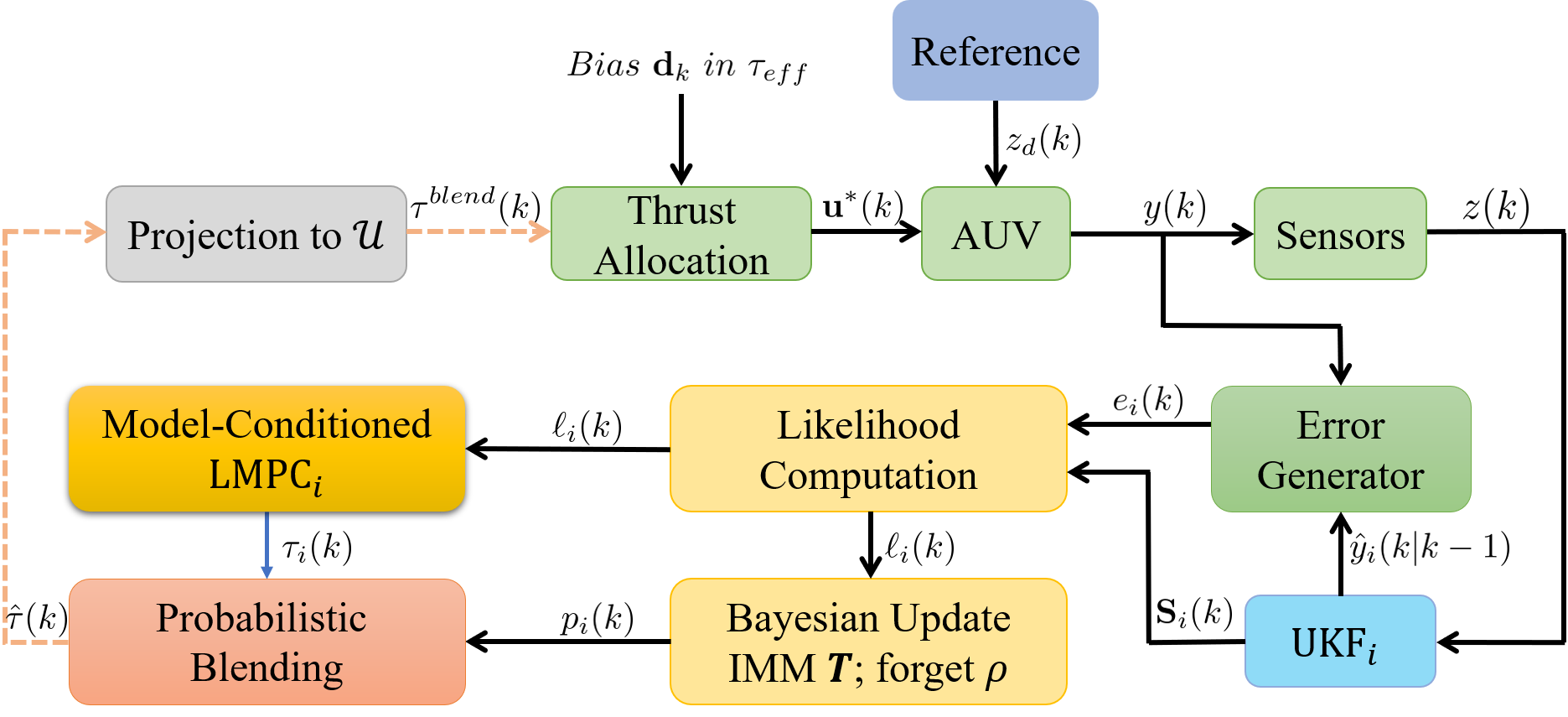}
    \caption{Bayesian multi-model LMPC architecture.Each model uses the same first-step Lyapunov contraction and amplitude/rate limits, with generalized forces $\bm\tau=\bm T(\bm\theta)\bm\Gamma\bm u$. An IMM Bayesian filter computes posterior probabilities for probability-weighted blending of the LMPC outputs.}
    \label{fig:ftc_overview}
\end{figure}

\Cref{fig:ftc_overview}  illustrates the overall control architecture, which integrates Bayesian mode estimation with LMPC to achieve fault-tolerant thrust allocation. A bank of Unscented Kalman Filters (UKFs) runs in parallel, each assuming a different thruster fault model.  At each time step a Bayesian estimator fuses the UKF outputs to compute posterior probabilities of each mode.  Each hypothesized mode is paired with its own LMPC controller, which is designed based on the corresponding fault model.  Crucially, all mode-conditioned LMPC controllers share a common Lyapunov function pair and a shared contraction constraint, so that recursive feasibility and closed-loop stability are guaranteed regardless of which controller is active.  Depending on the estimated mode probabilities, the system either selects the controller for the most likely mode (MAP selection) or forms a “soft” blend of the controllers’ outputs weighted by those probabilities.

This design is novel in combining Bayesian adaptation with LMPC for thrust control under faults.  By embedding thrust allocation directly into the LMPC optimization and enforcing a Lyapunov-inspired contraction constraint, the scheme inherits the stability properties of the base Lyapunov controller.  In effect, the MPC optimizer adjusts thrust commands online for best performance while respecting physical limits, and the contraction constraint ensures closed-loop stability.  This integrated approach is simpler and more flexible than architectures that cascade separate fault-tolerant loops.  In this approach, the controller passively adapts to each modeled failure mode without the need for explicit fault detection or reconfiguration. Building on the above architecture, we derive the mode-conditioned LMPC.

\subsection{Lyapunov-Constrained MPC with First-Step Contraction}\label{sec:LMPC}
We now integrate a Lyapunov-based constraint into MPC so that the optimal control inherits the stabilizing behavior of a nonlinear auxiliary controller while optimizing performance. The derivation follows the backstepping construction in~\cite{Shen_2018,Shen2018,Shen_2017,Wang2024,Gong2021}, extended here to the augmented bias state and online allocation.

Let the output position tracking error be $\tilde{\bm\eta}=\bm\eta-\bm\eta_d$. Define the virtual derivative reference and the \emph{velocity error}
\begin{equation}\label{eq:lmpc_vr_s}
    \dot{\bm\eta}_r = \dot{\bm\eta}_d - \tilde{\bm\eta},\qquad
    \bm s = \dot{\bm\eta}-\dot{\bm\eta}_r=\dot{\tilde{\bm\eta}}+\tilde{\bm\eta} .
\end{equation}
Using $\dot{\bm\eta}=\bm R(\psi)\bm v$, one has $\dot{\tilde{\bm\eta}}=\bm s-\tilde{\bm\eta}$. Let $\bm v_r=\bm R^\top(\psi)\dot{\bm\eta}_r$. Consider
\begin{equation}\label{eq:lmpc_V2}
    V_1=\tfrac12~\tilde{\bm\eta}^\top\bm K_p\tilde{\bm\eta},\quad
    V_2=\tfrac12~\bm s^\top \bm  M^{\ast}(\psi)\bm s+V_1,
\end{equation}
with $\bm K_p\!=\!\bm K_p^\top\!\succ0$ and $\bm  M^{\ast}(\psi)=\bm R(\psi)\bm M~\bm R^\top(\psi)$. Taking time derivative of $V_2$ and  Substituting the dynamics \eqref{eq:dynamics} results in\cite{Gong2021,Shi_2023,Shen2018}:
\begin{equation}\begin{aligned}\dot{V}_{2}=&-\mathbf{s}^{\top}[\mathbf{C}^{*}(\mathbf{v},\psi)+\mathbf{D}^{*}(\mathbf{v},\psi)]\mathbf{s}+\mathbf{s}^{\top}\mathbf{R}(\psi)\\&\times[\mathbf{\tau}-\mathbf{M}\dot{\mathbf{v}}_r-\mathbf{C}(\mathbf{v})\mathbf{v}_r-\mathbf{D}(\mathbf{v})\mathbf{v}_r]\\&+\frac{1}{2}\mathbf{s}^\top\dot{\mathbf{M}}^*(\psi)\mathbf{s}-\tilde{\eta}^\top\mathbf{K}_p\tilde{\eta}+\mathbf{s}^\top\mathbf{K}_p\tilde{\eta}\end{aligned}\end{equation}
where $\mathbf{C}^*(\mathbf{v},\psi)=\mathbf{R}(\psi)[\mathbf{C}(\mathbf{v})-\mathbf{M}\mathbf{R}^\top(\psi)\dot{\mathbf{R}}(\psi)]\mathbf{R}^\top(\psi)$ and $\mathbf{D}^*(\mathbf{v},\psi)=\mathbf{R}(\psi)\mathbf{D}(\mathbf{v})\mathbf{R}^\top(\psi)$. 

Because  $\bm C(\bm\nu)=-\bm C^\top(\bm\nu)$, we obtain
\begin{equation}\mathbf{s}^\top(\dot{\mathbf{M}}^*(\psi)-2\mathbf{C}^*(\mathbf{v},\psi))\mathbf{s}=0\quad\forall\;\mathbf{v},\;\psi\;,\mathbf{s}\end{equation}
Using standard Euler–Lagrange identities, the auxiliary  generalized-force law ~\cite{Shi_2023,ScienceDirect2020,Heshmati_Alamdari_2020}
\begin{equation}\label{eq:lmpc_tau_b}
    \bm\tau_b \!=\! \bm M\dot{\bm v}_r+\bm C(\bm v)\bm v_r+\bm D(\bm v)\bm v_r
    - \bm R^\top(\psi)\big(\bm K_p\tilde{\bm\eta}+\bm K_d\bm s\big),
\end{equation}
with $\bm K_d=\bm K_d^\top\!\succ0$, yields the decay
\begin{equation}\label{eq:lmpc_V2dot}
    \dot V_2 \;=\; -~\bm s^{\top}\!\big[\bm D^{*}(\bm v,\psi)+\bm K_d\big]\bm s \;-\; \tilde{\bm\eta}^\top \bm K_p \tilde{\bm\eta} \;\le\; 0,
\end{equation}

To capture slow-varying mismatch, we augment the generalized force with a bias $\bm d$ as in~\eqref{eq:tau_eff}, and define the composite Lyapunov function
\begin{equation}\label{eq:lmpc_V}
    V(\tilde{\bm\eta},\bm s,\bm d)=V_2(\tilde{\bm\eta},\bm s)+\tfrac12~\bm d^\top \bm P_d \bm d, \quad \bm P_d=\bm P_d^\top\succ0,
\end{equation}
where $\dot{\bm d}=-\bm\Lambda\bm d$ with $\bm\Lambda\succ0$ contributes $-\bm d^\top\bm P_d\bm\Lambda\bm d$ to $\dot V$.

Let $\bm\xi=[\tilde{\bm\eta}^ \top,\bm s^ \top,\bm d^ \top]^ \top$ and sampling period $\Delta t$. The MPC predictor uses the explicit Euler discretization\cite{Nikou_2020}
\begin{equation}\label{eq:lmpc_disc_pred}
    \bm\xi_{k+1} \;=\; \bm\xi_k + \Delta t~ \tilde{\bm f}(\bm\xi_k,\bm u_k;~\bm T(\bm\theta_k),\bm\Gamma_k),
\end{equation}
where $\tilde{\bm f}$ collects the error dynamics implied by~\eqref{eq:aug_error} and the allocation~\eqref{eq:tau_eff}.

Using $V$ as terminal penalty, the  Optimal Control Problem (OCP) at time $k$ is
\begin{align}
\min_{\{\bm u_{k+i}\}_{i=0}^{N-1}} \;\; & 
\sum_{i=0}^{N-1}\Big(\!\Vert \tilde{\bm\eta}_{k+i}\Vert_{\bm Q_\eta}^2 + \Vert \bm s_{k+i}\Vert_{\bm Q_s}^2 + \Vert \bm d_{k+i}\Vert_{\bm Q_d}^2 
\nonumber\\
&~+\Vert \Delta \bm u_{k+i}\Vert_{\bm R}^2\Big)+ V(\tilde{\bm\eta}_{k+N},\bm s_{k+N},\bm d_{k+N})
\label{eq:LMPC-discrete}
\\
\text{s.t.}\;\; &
\bm\xi_{k+i+1}=\bm\xi_{k+i}+\Delta t~\tilde{\bm f}(\bm\xi_{k+i},\bm u_{k+i}),\;\; i=0{:}N{-}1,\nonumber\\
& \bm\tau_{k+i}=\bm T(\bm\theta_k)\bm\Gamma_k~\bm u_{k+i},\nonumber\\
& \bm u_{\min}\le \bm u_{k+i}\le \bm u_{\max},\ \ \Vert \Delta \bm u_{k+i}\Vert_\infty\le \dot {\bm u}_{\max}\Delta t,\nonumber\\
& \underbrace{V(\bm\xi_{k+1})-V(\bm\xi_{k})\le -\alpha \Vert \tilde{\bm\eta}_{k}\Vert^2}_{\text{Lyapunov descent at the first step}},\ \alpha>0,\nonumber\\
& \bm\xi_{k+N}\in \mathcal{X}_f=\{\bm\xi: V(\bm\xi)\le c\}.\nonumber
\end{align}

\noindent\emph{Terminal set.} In this work we choose a small $c>0$ so that $\mathcal X_f=\{\boldsymbol{\xi}:V(\boldsymbol{\xi})\le c\}$ lies within the local region of attraction(ROA) of the auxiliary backstepping feedback and respects input/rate constraints; this set is obtained numerically.
The \emph{first-step} descent inequality enforces that the MPC’s initial move decreases $V$ at least by $\alpha\Vert \tilde{\bm\eta}_k\Vert^2$, matching the decay of the auxiliary law~\eqref{eq:lmpc_tau_b}. The terminal set $\mathcal X_f$ is chosen positively invariant under~\eqref{eq:lmpc_tau_b} with the \emph{faulty} mapping $\bm T(\bm\theta)\bm\Gamma$.

\subsection{Online Reconfiguration of Thrust Effectiveness}
We model thruster faults by \emph{column-wise scaling and in-plane rotation} of the thrust mapping (see Sec.~\ref{section:PRELIMINARIES} for the parameterization).
Let $\bm T=[\bm t_1~\bm t_2~\bm t_3~\bm t_4]\in\mathbb{R}^{3\times4}$ be the nominal allocation and
$\bm\Gamma=\operatorname{diag}(\gamma_1,\ldots,\gamma_4)$ with $\gamma_i\in[0,1]$.
If thruster $i$ is misaligned by $\theta_i$ and derated, we form the rotated mapping
\begin{equation}\label{eq:allocation-matrix}
  \tilde{\bm T}(\bm\theta)
  =\big[\bm t_1,\ldots, \bm R(\theta_i)\bm t_i,\ldots,\bm t_4\big],\;
  \bm\tau_k=\tilde{\bm T}(\bm\theta_k)\,\bm\Gamma_k\,\bm u_k ,
\end{equation}
where $\bm R(\theta_i)$ denotes the planar rotation applied to the $i$th column.

At each sampling instant $k$, the OCP in \eqref{eq:LMPC-discrete} is updated with the current
$\tilde{\bm T}(\bm\theta_k)$ and $\bm\Gamma_k$.
Per-thruster bounds are tightened to reflect the remaining capability; in particular, if $\gamma_i=0$ then $u_{i,k}=0$.
Feasibility is enforced by the amplitude and rate limits in \eqref{eq:LMPC-discrete}.

We warm-start the solver with a least-squares allocation
\begin{equation}
  \bm u_k^{\mathrm{init}}
  = \big(\tilde{\bm T}(\bm\theta_k)\,\bm\Gamma_k\big)^{\dagger}\,\bm\tau_k ,
\end{equation}
or, for numerical robustness, with a damped least-squares form
\begin{equation}
  \bm u_k^{\mathrm{init}}
  = \big(\tilde{\bm T}(\bm\theta_k)\,\bm\Gamma_k\big)^{\!\top}
    \Big(\tilde{\bm T}(\bm\theta_k)\,\bm\Gamma_k\big(\tilde{\bm T}(\bm\theta_k)\,\bm\Gamma_k\big)^{\!\top}
          + \varepsilon \bm I \Big)^{-1}\bm\tau_k ,
          \nonumber
  \end{equation}

Where $\varepsilon>0$. Because the first-step Lyapunov contraction in \eqref{eq:LMPC-discrete} is evaluated using the \emph{current} mapping
$\tilde{\bm T}(\bm\theta_k),\bm\Gamma_k$, reconfiguration does not alter the contraction condition.
It only changes the admissible input set and the predicted generalized force, while the enforced decrease of $V$ at the first step remains in force.

\section{Bayesian Multi-Model Adaptive Fault-Tolerant Framework}\label{sec:bayes}
After establishing the single-model LMPC with a common Lyapunov contraction constraint, we now describe the Bayesian multi-model adaptation layer that provides online fault diagnosis and controller blending. The key idea is to run a bank of mode-conditioned estimators and LMPCs, update the posterior probabilities of the modes from measurement innovations, and fuse the candidate control actions using these posteriors to achieve \emph{soft switching}  with stability guarantees.

\subsection{Multi-Model Parameterization }
We adopt a discrete set of mode hypotheses
\begin{equation}
  \mathcal M = \{I, II, III\},
\end{equation}
where each mode $i\in\mathcal{M}$ is characterized by the thruster effectiveness and possible misalignment parameters
\begin{equation}
\Theta_i=\{\Gamma_i,\theta_i\},\;
\Gamma_i=\mathrm{diag}(\gamma_{1,i},\ldots,\gamma_{4,i})\in[0,1]^{4\times4},    
\end{equation}
and the column-wise rotated allocation matrix are described in \eqref{eq:allocation-matrix}.
The plant dynamics in mode $i$ keep the \emph{same} nonlinear structure as in Sec.~\ref{section:PRELIMINARIES}, changing only through the generalized force mapping
\begin{equation}
   \bm \tau = \tilde{\bm T}(\theta_i) \bm{\Gamma}_i~\bm u,\qquad
 \bm\tau_{\mathrm{eff}}=\bm \tau+\bm d, 
\end{equation}
with the same augmented bias dynamics $\dot {\bm d}=- \bm \Lambda \bm d$. Throughout this section, the LMPC layer per mode $i$ uses the identical cost and constraints as Sec.~\ref{sec:LMPC}.

\subsection{Bayesian Posterior Probability Update}\label{sec:bayes_update}

Each mode $i$ runs a UKF that provides the one-step predicted output $\hat{y}_i(k|k-1)$ and innovation covariance $S_i(k)$. We update the posterior probabilities $p_i(k)$ via an IMM prior and a Gaussian log-likelihood, and we use probability-weighted blending of mode-conditioned LMPC outputs\cite{Julier2004} . Let the innovation be 
\begin{equation}
   e_i(k)=y(k)-\hat y_i(k|k-1). 
\end{equation}
 
Assuming Gaussian measurement noise, the instantaneous log-likelihood \cite{Jazwinski2007,Blom_1988} is computed as
\begin{equation}\label{eq:loglike}
\ell_i(k) = -\tfrac{1}{2}\!\left[e_i^\top S_i^{-\!1} e_i + \log\det S_i + p\log(2\pi)\right],
\end{equation}
where $p=\dim(y(k))$. To prevent numerical singularity when $S_i$ is ill-conditioned, a small $\bm \varepsilon>0$ is added to the diagonal, i.e., $S_i\leftarrow S_i+\bm \varepsilon I$.

To suppress chattering in probability evolution caused by transient noise, we apply an exponentially weighted cumulative log-likelihood \cite{Simon_2006}:
\begin{equation}
\bar\ell_i(k)=\rho~\bar\ell_i(k-1)+\ell_i(k),\quad \rho\in(0,1],
\end{equation}
where $\rho$ is a forgetting factor. When $\rho=1$, all past likelihoods are equally weighted; smaller $\rho$ values discount older measurements.

Slow mode transitions are modeled by a first-order Markov chain with transition matrix $T_{ji}=\Pr\{m(k)=i\mid m(k-1)=j\}$, which yields the mixed prior probability \cite{Blom_1988,Ackerson1970,Bar_Shalom_2002}:
\begin{equation}
\tilde p_i(k) = \sum_{j\in\mathcal M} T_{ji}~p_j(k-1).
\end{equation}
This is equivalent to the mode interaction step in the Interacting Multiple Model (IMM) filter \cite{X_Rong_Li_2005}.

Applying Bayes’ theorem, the posterior probability is
\begin{equation}\label{eq:bayes_raw}
p_i(k)=\frac{\tilde p_i(k)~\exp\{\bar\ell_i(k)\}}{\sum_{r\in\mathcal M} \tilde p_r(k)~\exp\{\bar\ell_r(k)\}},
\quad \sum_{i} p_i(k)=1.
\end{equation}
For improved numerical stability, especially when $\bar\ell_i(k)$ is large in magnitude, \eqref{eq:bayes_raw} is computed in the log domain using the log-sum-exp technique \cite{Ristic2003}:
\begin{align}
\hat\ell_i(k) &= \log\tilde p_i(k) + \bar\ell_i(k),\;\;
m = \max_{r\in\mathcal M} \hat\ell_r(k),\\
p_i(k) &= \frac{\exp\{\hat\ell_i(k)-m\}}{\sum_{r\in\mathcal M} \exp\{\hat\ell_r(k)-m\}}.
\end{align}
This approach preserves probability normalization and avoids underflow or overflow in floating-point computation.

\subsection{Control Fusion and Supervisory Logic}\label{sec:fusion}
At each sampling instant, the mode–conditioned LMPC for $i\in\mathcal M$ returns a feasible input $\bm u_i^\star(k)$ and the corresponding generalized force $\bm\tau_i^\star(k)=\bm T(\theta_i)\bm\Gamma_i\bm u_i^\star(k)$. We apply a \emph{probabilistic blending} of these candidates which form the convex combination in force space~\cite{LAINIOTIS1972,Lainiotis1974,Deshpande1973}
\begin{equation}\label{eq:blend}
\bm \tau^{\mathrm{blend}}(k)=\sum_{i\in\mathcal{M}} p_i(k)~ \bm \tau_i^\star(k),
\end{equation}
and map it to thruster commands via a least–squares allocation using the current most–probable parameters $(\bar\theta,\bar\Gamma)=\arg\max_i p_i(k)$:
\begin{equation}
\hat{\bm u}(k)=\big(\bm T(\bar\theta)\bar{\bm\Gamma}\big)^\dagger~\bm\tau^{\mathrm{blend}}(k),\;
\bm u^{\mathrm{blend}}(k)=\Pi_{\mathcal U}\!\big(\hat{\bm u}(k)\big),
\end{equation}
where $\Pi_{\mathcal U}$ denotes projection onto the admissible input set.
This procedure yields smooth control during probability transitions and  guarantees feasibility by construction through $\Pi_{\mathcal U}$.

For each mode $i$, the associated LMPC enforces the same first–step contraction
\begin{equation}\label{eq:first_step_contraction}
V\!\left(F(\bm\xi_k,\bm u_i^\star)\right)-V(\bm\xi_k)\le -\alpha \|\tilde{\bm x}_k\|^2,\quad \alpha>0.
\end{equation}
Over one prediction step, the augmented error dynamics are affine in the generalized force   and $V(\cdot)$ is quadratic; hence for the unprojected blend $\hat{\bm u}$ one has the Jensen-type bound
\begin{equation}\label{eq:jensen}
V\!\left(F(\bm\xi_k,\hat{\bm u})\right)\;\le\; \sum_{i\in\mathcal M} p_i(k)~V\!\left(F(\bm\xi_k,\bm u_i^\star)\right).
\end{equation}
Combining \eqref{eq:first_step_contraction}–\eqref{eq:jensen} and $\sum_i p_i=1$ yields
\begin{equation}\label{eq:blend_contract}
V\!\left(F(\bm\xi_k,\hat{\bm u})\right)-V(\bm\xi_k)\;\le\; -\alpha \|\tilde{\bm x}_k\|^2.
\end{equation}
If $\hat{\bm u}$ is feasible, the applied $\bm u^{\mathrm{blend}}=\hat{\bm u}$ preserves the exact contraction \eqref{eq:blend_contract}. If projection is active, Lipschitz continuity of $V\!\circ F$ in $\bm u$ implies
\begin{equation}
V\!\left(F(\bm\xi_k,\bm u^{\mathrm{blend}})\right)-V(\bm\xi_k)
\;\le\; -\alpha \|\tilde{\bm x}_k\|^2 + L_u~\|\bm u^{\mathrm{blend}}-\hat{\bm u}\|,
\end{equation}
for some $L_u>0$ depending on local model bounds. Since $\Pi_{\mathcal U}$ is a nonexpansive projection and rate limits bound $\|\bm u^{\mathrm{blend}}-\hat{\bm u}\|$, the contraction margin remains negative provided $L_u~\|\bm u^{\mathrm{blend}}-\hat{\bm u}\|\le \alpha'\|\tilde{\bm x}_k\|^2$ for a chosen $\alpha'\in(0,\alpha)$. In practice, this is ensured by  the shared auxiliary backstepping rate embedded in \eqref{eq:first_step_contraction}, and  conservative rate limits, so recursive feasibility and closed-loop stability are maintained during probability evolution.
To reduce computation without affecting stability, one may temporarily switch to the most–probable controller when $p_m(k)\ge p_{\mathrm{on}}$ for $N_{\mathrm{on}}$ samples, and resume blending only if $p_m(k)\le p_{\mathrm{off}}<p_{\mathrm{on}}$ for $N_{\mathrm{off}}$ samples. Anomaly-triggered “unlocking’’  can be used to avoid probability lock-in under new faults.

\section{Simulations and Results}\label{simulation_results}
To demonstrate the effectiveness and fault-tolerant capability of the proposed ALMPC framework, we conduct numerical simulations on the AUV model. All simulations are implemented in MATLAB R2023b. Model and controller parameters follow the Saab Seaeye Falcon specifications \cite{Shen2018,Shen_2018,Shi_2023}.

\subsection{Simulation Settings}
Three representative AUV modes are considered:
\begin{itemize}
    \item \textbf{Model\; I (nominal)} $\gamma_i=1,\theta_i=0$; 
    \item \textbf{Model\; II (thruster~\#1 failure)} $\gamma_{1}=0$;
    \item  \textbf{III (thruster~\#3 misaligned + derated)} $\gamma_{3}=0.3$, column $t_3$ rotated by $\theta_{3} =15^{\circ}$.
\end{itemize}
This parameterization directly matches the thrust mapping in Sec.~\ref{section:PRELIMINARIES} and the LMPC constraints in Sec.~\ref{sec:LMPC}.
Two scenarios are evaluated:
\textbf{Case I}: Model~I $\rightarrow$ Model~II at $t=15~\mathrm{s}$. The reference trajectory is
\[
p(t)=\begin{cases}
x_d=0.5~t,\\
y_d=\sin(0.5~t).
\end{cases}
\]
\textbf{Case II}: Model~I $\rightarrow$ Model~II at $t=10~\mathrm{s}$, then Model~II $\rightarrow$ Model~III at $t=20~\mathrm{s}$. The reference is an eight-shaped path
\[
p(t)=\begin{cases}
x_d=-\sin(0.5~t),\\
y_d=\sin(0.25~t).
\end{cases}
\]

Unless otherwise stated, the controller uses sampling time $\Delta t=0.1~\mathrm{s}$ and prediction horizon $N=10$. The LMPC weights follow ~\cite{Shen_2018}
\begin{equation}
\begin{aligned}
  Q&=\mathrm{diag}(10^{5},10^{5},10^{3},10^{2},10^{2},10^{2})\\
R&=\mathrm{diag}(10^{-4},10^{-4},10^{-4},10^{-4})\\
Q_f&=\mathrm{diag}(10^{3},10^{3},10^{2},10,10,10).  
\end{aligned}  
\nonumber
\end{equation}

The UKF covariances are
\begin{equation}
    \begin{aligned}
           Q_{\text{ukf}}&=\mathrm{blkdiag}(0.01,0.01,0.01,~0.005,0.005,0.005)\\
R_{\text{ukf}}&=\mathrm{diag}(0.1,0.1,0.1,~0.03,0.03,0.03).  
    \end{aligned}
    \nonumber
  \end{equation}

Thruster amplitude limits are $\pm 500~\mathrm{N}$. The initial state is $x(0)=[0.5,~0,~0,~0,~0,~0]^\top$. The baseline BSC gains are $K_p=K_d=\mathrm{diag}(1,1,1)$. The LMPC problem in \eqref{eq:LMPC-discrete} is solved by discretizing the dynamics and applying SQP to the resulting Karush–Kuhn–Tucker (KKT) system \cite{Antoniou_2021}. Additionally, the Bayesian layer uses a first-order Markov prior with diagonal persistence \(T_{\mathrm{diag}}=0.98\), i.e., \(\bm T=T_{\mathrm{diag}}\bm I_{3}+\tfrac{1-T_{\mathrm{diag}}}{2}(\bm 1\bm 1^\top-\bm I_{3})\); the log-likelihood is exponentially weighted with \(\rho=0.9995\). 
To stress-test offset rejection we set the bias leakage in \eqref{eq:tau_eff} to \(\bm\Lambda=\bm 0\). 
Supervisory switching uses hysteresis thresholds \(p_{\mathrm{on}}=0.95,~p_{\mathrm{off}}=0.80\) and dwell counts \(N_{\mathrm{on}}=10,~N_{\mathrm{off}}=5\).
Detection time $T_{\text{det}}$ is measured from the fault instant to the earliest time when the true-mode posterior exceeds $p_{\mathrm{on}}=0.95$ for $N_{\mathrm{on}}=10$ consecutive samples; accommodation time $T_{\text{acc}}$ is measured to the earliest time when $|e_x|,|e_y|<0.1~\mathrm{m}$ for at least $1~\mathrm{s}$.

\subsection{ Tracking Performance}
\begin{figure}
    \centering
    \includegraphics[width=1\columnwidth]{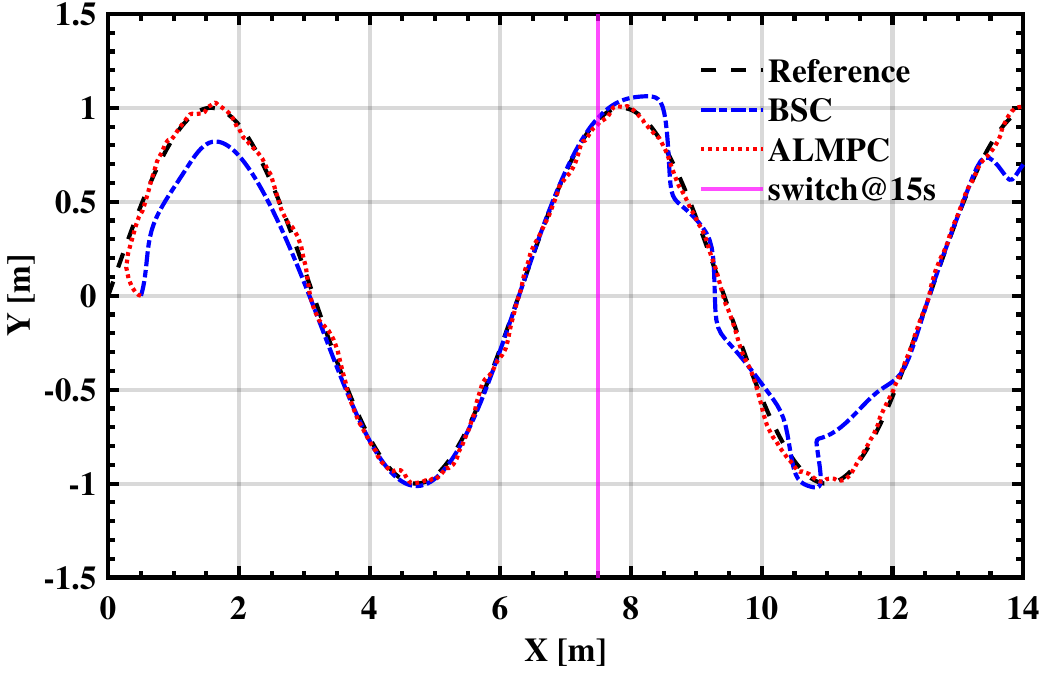}
    \caption{AUV trajectory in the XY plane under single-model switching.}
    \label{fig:1to2 tracking results}
\end{figure}

\begin{figure}
    \centering
    \includegraphics[width=1\columnwidth]{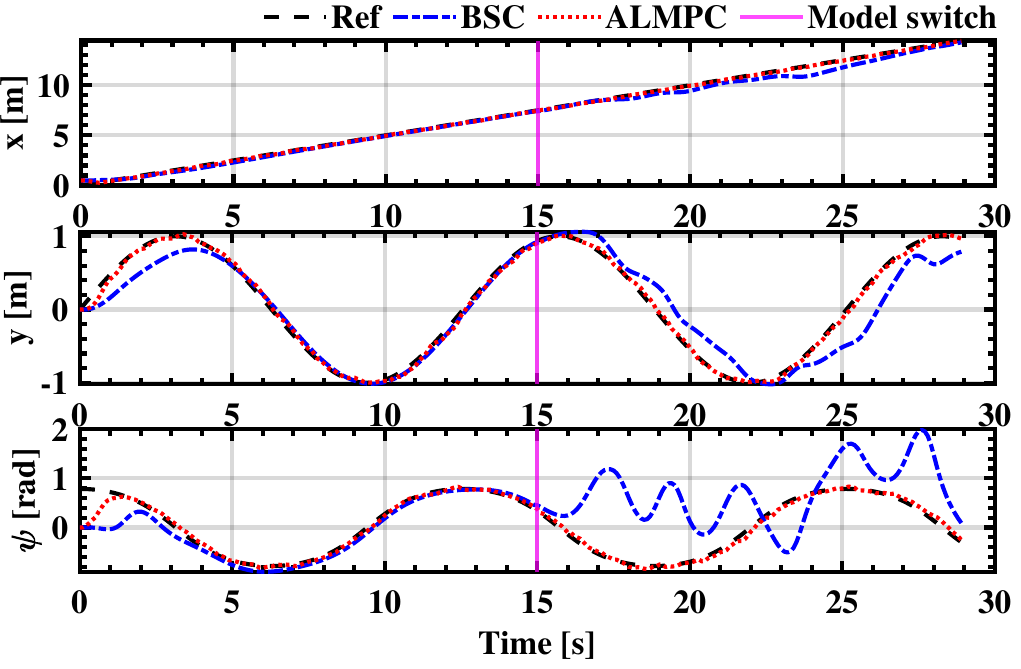}
    \caption{State trajectories $(x,y,\psi)$ for single-model switching.}
    \label{fig:1to2 tracking Performance}
\end{figure}

Results for \textbf{Case I} are shown in \Cref{fig:1to2 tracking results,fig:1to2 tracking Performance}. The black solid curve is the reference, the blue dash-dot curve is BSC, and the red dashed curve is ALMPC. The vertical magenta line marks the switch at $t=15~\mathrm{s}$.

\begin{figure}
    \centering
    \includegraphics[width=1\columnwidth]{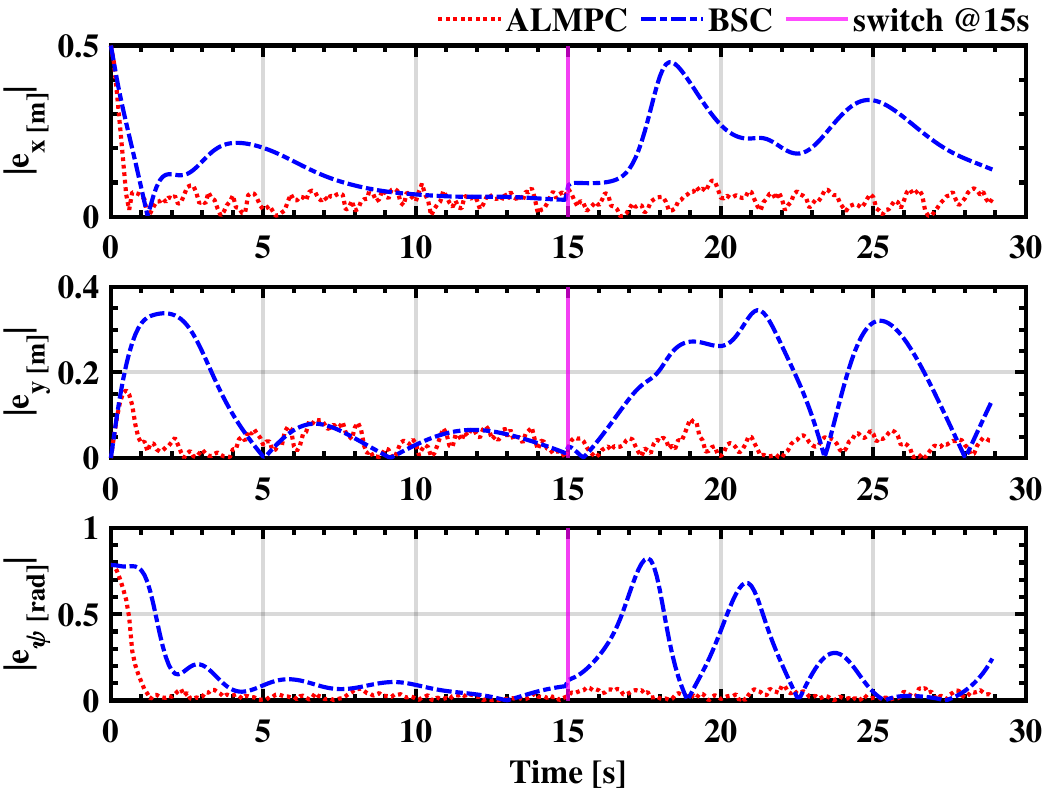}
    \caption{Absolute tracking errors under single-model switching.}
    \label{fig:1to2 absolute tracking error}
\end{figure}

Before the fault, both BSC and ALMPC track well. After thruster \#1 fails, BSC degrades markedly due to loss of control authority, whereas ALMPC swiftly reallocates thrust among healthy thrusters and preserves tight tracking; see the absolute errors in \Cref{fig:1to2 absolute tracking error}. The x-axis error of BSC peaks near $1.0~\mathrm{m}$ around $t=24~\mathrm{s}$, while ALMPC remains below $0.1~\mathrm{m}$; the y-axis error shows a similar gap (BSC up to $0.8~\mathrm{m}$ vs. ALMPC $<0.05~\mathrm{m}$).

\begin{figure}
    \centering
    \includegraphics[width=1\columnwidth]{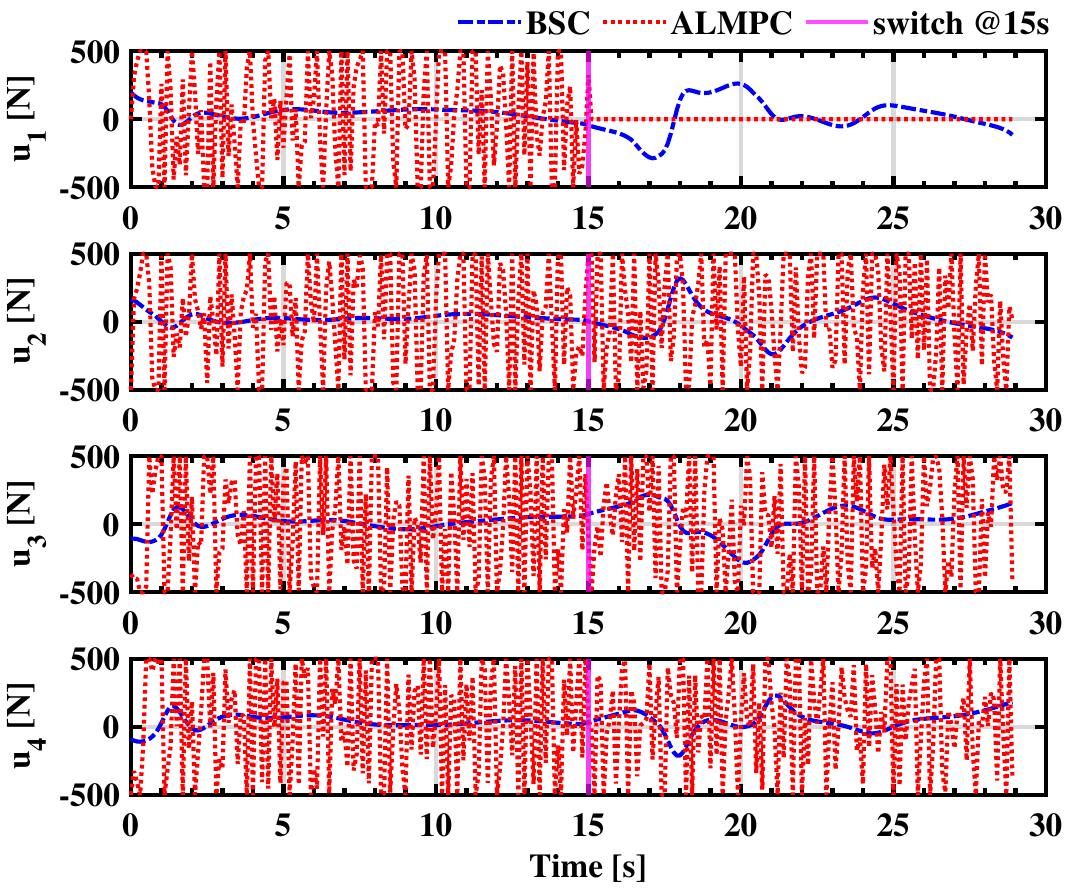}
    \caption{Thruster outputs under single-model switching.}
    \label{fig:1to2 Thruster Output}
\end{figure}

Thruster commands in \Cref{fig:1to2 Thruster Output} highlight the effect of reallocation: BSC continues to request force from the failed thruster \#1, wasting effort and distorting the trajectory, whereas ALMPC shifts demand to the remaining thrusters while respecting limits.

\begin{figure}
    \centering
    \includegraphics[width=1\columnwidth]{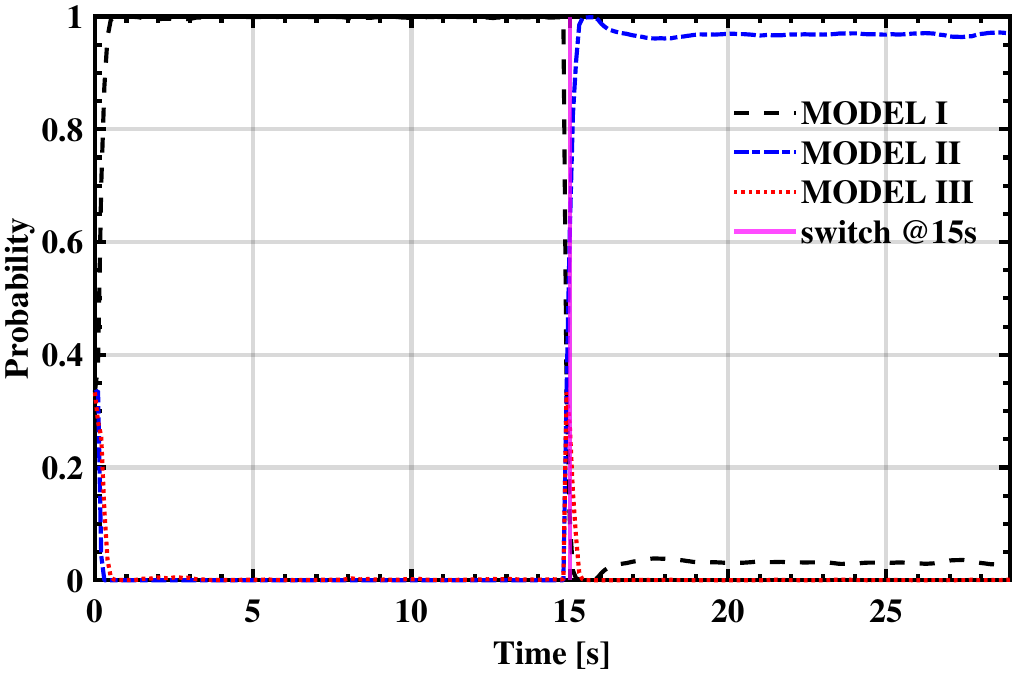}
    \caption{Model posterior probabilities for single-model switching.}
    \label{fig:1to2 probabilities}
\end{figure}

The Bayesian identification is shown in \Cref{fig:1to2 probabilities}. The posterior, initially dominated by Model~I, rapidly shifts after $t=15~\mathrm{s}$ and exceeds $0.9$ for Model~II in about $1~\mathrm{s}$, while the other probabilities decay toward zero. During this transient, probability-weighted blending keeps the control smooth and prevents switching kicks. Using the supervisor’s confirmation rule $(p_{\mathrm{on}}=0.95,~N_{\mathrm{on}}=10)$, the detection delay is
$T_{\mathrm{det}}\approx~$\textit{0.3}$~\mathrm{s}$.
With the error band $(\bm \varepsilon_\eta,\bm \varepsilon_\psi)=(0.10~\mathrm{m},~0.05~\mathrm{rad})$,
the accommodation time is $T_{\mathrm{acc}}\approx~$\textit{0.3}$~\mathrm{s}$.

\subsection{Comparative Evaluation}
\begin{figure}
    \centering
    \includegraphics[width=1\columnwidth]{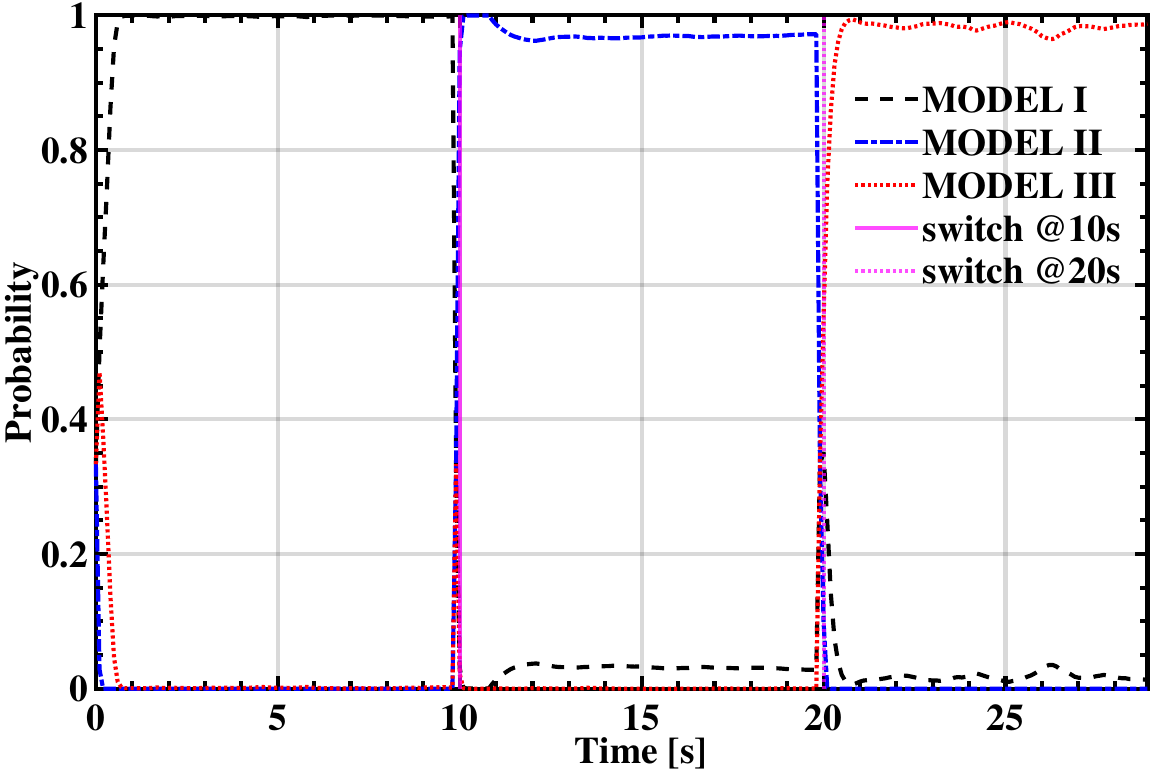}
    \caption{Model posterior probabilities under multiple-model switching.}
    \label{fig:twice switch probabilities}
\end{figure}

We further compare {ALMPC}, { (AMPC)}, and {BSC} under \textbf{Case II} . All estimation settings are identical to Case~I. The posteriors in \Cref{fig:twice switch probabilities} confirm rapid recognition: after the failure at $t=10~\mathrm{s}$, the probability of Model~II exceeds  $0.95$ within $0.3~\mathrm{s}$ (first switch) and $0.5~\mathrm{s}$ (second switch) under the confirmation rule; following the second fault at $t=20~\mathrm{s}$, Model~III becomes dominant with a similar time-to-confidence.
For the first transition (I$\!\to$II at $t{=}10$~s), the confirmation time is
$T_{\mathrm{det}}\approx~$\textit{0.3}$~\mathrm{s}$ and the accommodation time
$T_{\mathrm{acc}}\approx~$\textit{0.3}$~\mathrm{s}$.
For the second transition (II$\!\to$III at $t{=}20$~s), we obtain
$T_{\mathrm{det}}\approx~$\textit{0.5}$~\mathrm{s}$ and
$T_{\mathrm{acc}}\approx~$\textit{0.4}$~\mathrm{s}$ under the same criteria.

\begin{figure}
    \centering
    \includegraphics[width=1\columnwidth]{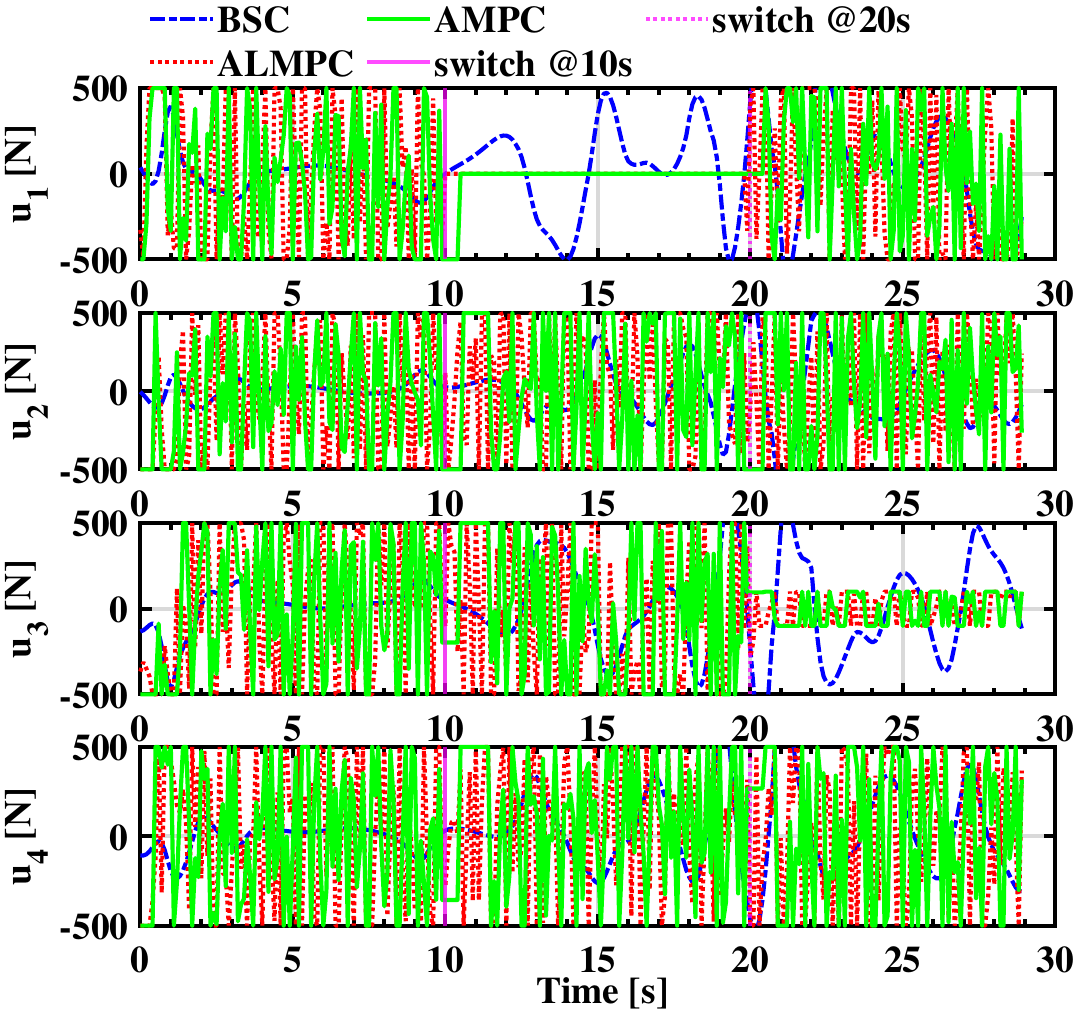}
    \caption{Thruster outputs under multiple-model switching.}
    \label{fig:twice  Thruster Output}
\end{figure}

The thruster responses in \Cref{fig:twice  Thruster Output} make the reallocation explicit: BSC continues to command the failed thruster, while ALMPC increases the contribution of healthy thrusters (e.g., thruster \#2) immediately after the failures.

\begin{figure}
    \centering
    \includegraphics[width=1\columnwidth]{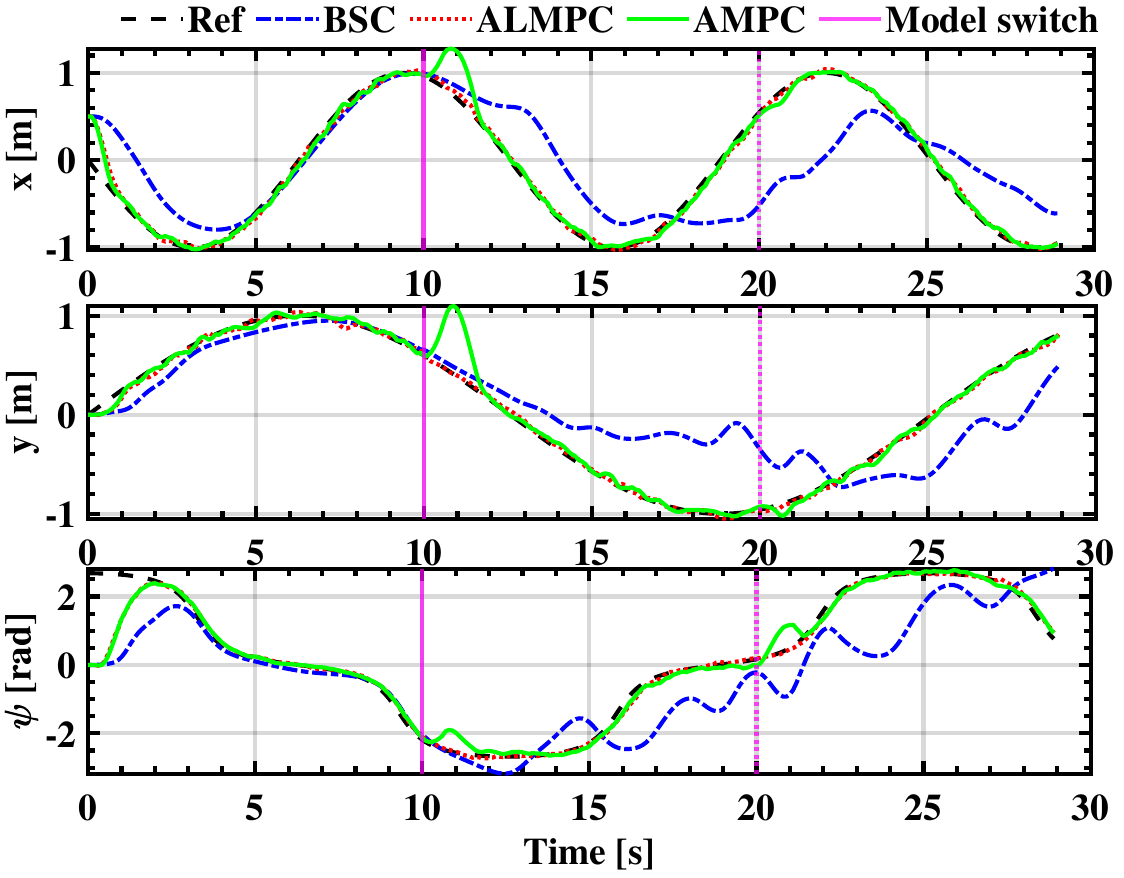}
    \caption{State trajectories $(x,y,\psi)$ under multiple-model switching.}
    \label{fig:twice  switch tracking Performance}
\end{figure}

\begin{figure}
    \centering
    \includegraphics[width=1\columnwidth]{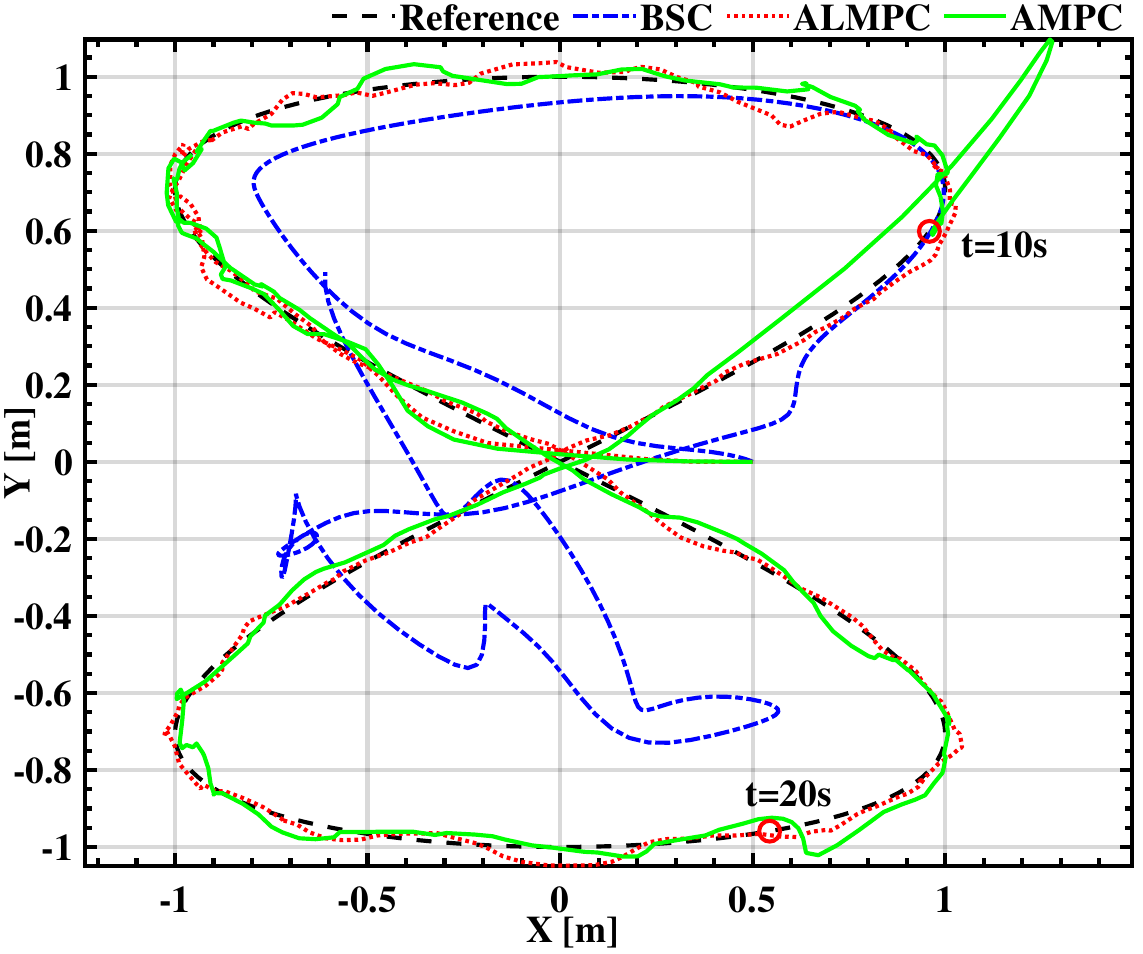}
    \caption{AUV trajectory in the XY plane under multiple-model switching.}
    \label{fig:twice  switch tracking results}
\end{figure}

\begin{figure}
    \centering
   \includegraphics[width=1\columnwidth]{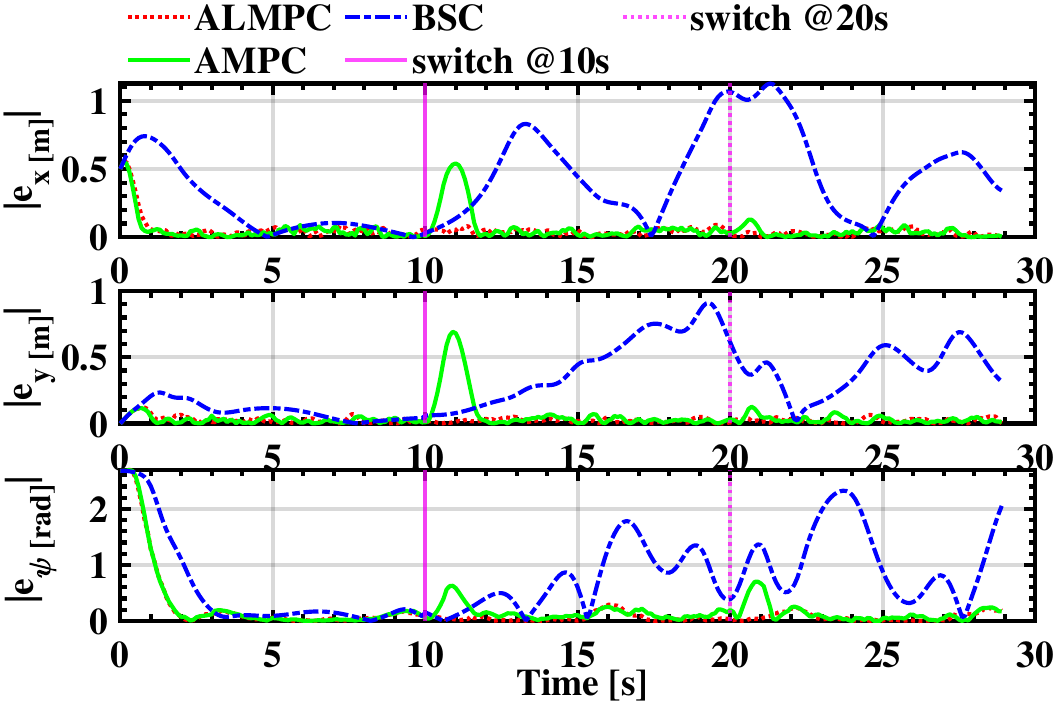}
    \caption{Absolute tracking errors for different control methods.}
    \label{fig:twice  switch tracking error}
\end{figure}

Trajectory and state comparisons are given in \Cref{fig:twice  switch tracking results,fig:twice  switch tracking Performance}. Prior to $10~\mathrm{s}$, ALMPC and AMPC perform similarly, while BSC lags. At the fault instants ($10~\mathrm{s}$ and $20~\mathrm{s}$), \emph{AMPC exhibits pronounced transients}, whereas \emph{ALMPC remains smooth}. The absolute errors in \Cref{fig:twice  switch tracking error} show that ALMPC keeps deviations within $0.1~\mathrm{m}$ in both $x$ and $y$ even during fault transitions, whereas AMPC peaks near $0.5~\mathrm{m}$ and BSC exceeds $1~\mathrm{m}$.

At a fault instant, AMPC’s internal model $\hat\Theta$ lags the true post-fault parameters $\Theta^\star$; hence the first control move is computed against a mismatched prediction/constraint set. Because AMPC does not enforce a first-step Lyapunov contraction, $V(\xi_{k+1})-V(\xi_k)$ may temporarily increase, manifesting as an overshoot and a sharp change in thrust. In contrast, each mode-conditioned LMPC in the proposed framework satisfies the \emph{same} contraction inequality, and the applied input is the probability-weighted blend $\bm u=\sum_i p_i \bm u_i^{*}$, projected onto the admissible set. Since $\sum_i p_i=1$, the convex combination preserves contraction, thus preventing switching kicks and keeping thrust continuous up to the enforced rate limits.
\begin{table}[t]
\centering
\caption{Quantitative tracking metrics under multiple-model switching (Case~II). $x,y$ in $m$; $\psi$ in $rad$. ; IAE is $\int |e|~dt$.}
\label{tab:quant_metrics}
\small
\setlength{\tabcolsep}{4pt}
\begin{tabular}{l l c c c}
\toprule
State & Metric & ALMPC & AMPC & BSC \\
\midrule
$x$ & RMSE   & 0.0799 & 0.1415 & 0.2704 \\
    & MaxAE  & 0.5501 & 0.6283 & 0.7402 \\
    & SSE    & 1.8531 & 5.8072 & 21.197 \\
    & IAE    & 1.2399 & 2.1132 & 5.8113 \\
\midrule
$y$ & RMSE   & 0.0296 & 0.1072 & 0.1355 \\
    & MaxAE  & 0.1267 & 0.4857 & 0.3011 \\
    & SSE    & 0.2543 & 3.3333 & 5.3224 \\
    & IAE    & 0.6226 & 1.6552 & 3.2003 \\
\midrule
$\psi$ & RMSE & 0.4802 & 0.5646 & 0.6803 \\
       & MaxAE& 2.6960 & 2.7014 & 2.6779 \\
       & SSE  & 66.857 & 92.444 & 134.200 \\
       & IAE  & 4.8767 & 7.8114 & 11.171 \\
\bottomrule
\end{tabular}
\end{table}
Relative to AMPC, ALMPC reduces RMSE by $44\%$ (x), $72\%$ (y), and $15\%$ ($\psi$), and cuts IAE by $41$--$62\%$ across all states. 

As summarized in Table~\ref{tab:quant_metrics}, ALMPC attains the lowest RMSE and MaxAE across all components while keeping the squared error (SSE) small, in agreement with the smooth, probability-weighted blending across Lyapunov-Constrained LMPCs. By contrast, AMPC’s transients at the fault instants are consistent with the parameter lag and the absence of a first-step contraction bound.
\section{Conclusion}
This work presented an ALMPC with a Bayesian multi-model layer for fault-tolerant AUV tracking. The design meets the three stated objectives: (i) high-accuracy nominal tracking via an augmented error–bias model that removes steady-state offsets; (ii) fast fault detection and accommodation through a UKF-driven Bayesian update and probabilistic blending—quantitatively, in Case~II we obtained $T_{\text{det}}{=}\{0.3,~0.5\}~\mathrm{s}$ and $T_{\text{acc}}{=}\{0.3,~0.4\}~\mathrm{s}$ at the two switches; and (iii) stable thrust reallocation under degradation guaranteed by a shared first-step Lyapunov contraction across modes. Compared with AMPC and BSC, ALMPC delivered the lowest RMSE/MaxAE and significantly smaller SSE/IAE (Table~\ref{tab:quant_metrics}) while keeping commands smooth at mode transitions.

Future work will extend the mode set to continuous fault manifolds, develop tighter robustness margins for rapidly varying posteriors, and validate the approach in 6-DOF experiments and sea trials.


\bibliographystyle{Bibliography/IEEEtranTIE}
\bibliography{Bibliography/LMPC-ACLSS}\ 

\end{document}